\documentclass[superscriptaddress,floatfix,twocolumn,prl]{revtex4}

\usepackage{amsmath}
\usepackage[dvips]{graphicx}
\usepackage{epsfig}
\usepackage{tabularx}
\usepackage{color}
\usepackage[pdfpagemode=UseNone,colorlinks=true,linkcolor=blue,citecolor=blue]{hyperref}

\usepackage{soul}
\usepackage{gensymb}

\begin{document}

\title[]{An ultra-narrow linewidth levitated nano-oscillator for testing dissipative wavefunction collapse }

\author{A. Pontin}
 \email{a.pontin@ucl.ac.uk}
\affiliation{Department of Physics \& Astronomy, University College London, Gower Street, London WC1E 6BT, United Kingdom}

\author{N. P. Bullier}%
\affiliation{Department of Physics \& Astronomy, University College London, Gower Street, London WC1E 6BT, United Kingdom}%

\author{M. Toro\v{s}}%
\affiliation{Department of Physics \& Astronomy, University College London, Gower Street, London WC1E 6BT, United Kingdom}%

\author{P. F. Barker}
\email{p.barker@ucl.ac.uk}
\affiliation{Department of Physics \& Astronomy, University College London, Gower Street, London WC1E 6BT, United Kingdom}%

\maketitle

\textbf{Levitated nano-oscillators are seen as promising platforms for testing fundamental physics and testing quantum mechanics in a new high mass regime~\cite{PhysRevLett.107.020405,PhysRevLett.105.101101, PhysRevLett.110.071105,PhysRevLett.113.251801,Bateman1}. Levitation allows extreme isolation from the environment, reducing the decoherence processes that are crucial for these sensitive experiments~\cite{quidant_recoil,giacomo1,Li2011,gieseler1}. A fundamental property of any oscillator is its linewidth and  mechanical quality factor, Q. Narrow linewidths in the microHertz regime and mechanical Q's as high as 10$^{12}$ have been predicted for levitated systems ~\cite{ChangPNAS}, but to date, the poor stability of these oscillators over long periods have prevented direct measurement in high vacuum. Here we report on the measurement of an ultra-narrow linewidth levitated nano-oscillator, whose width of $81\pm\,23\,\mu$Hz is only limited by residual gas pressure at high vacuum. This narrow linewidth allows us to put new experimental bounds on dissipative models of wavefunction collapse including continuous spontaneous localisation and Di\'{o}si-Penrose~\cite{PRA_dCSL,CSL1,CSL2,Smirne2015,dDP,Penrose1996,diosi} and illustrates its utility for future precision experiments that aim to test the macroscopic limits of quantum mechanics~\cite{goldwater,PRA_TEQ}}.

Levitated oscillators formed by trapping neutral and charged nanoparticles in optical~\cite{barker2010,ChangPNAS,oriol_2010}, electric~\cite{giacomo1,giacomo2} or magnetic fields~\cite{oriol_magnetic,Slezak2018} are unlike any other optomechanical platform in that the oscillator's properties can be widely tuned by control of the levitating fields. The fields can even be turned-off, offering low noise, field free, measurements for short periods of time~\cite{PhysRevLett.121.063602,Bateman1}. A fundamental property of any oscillator is its linewidth and mechanical quality factor, Q. Narrow linewidths in the sub-microHertz range and quality factors has high as 10$^{12}$ have been predicted for optical levitated systems. However, the poor stability of these oscillators over long periods, coupled with their tendency to operate in anharmonic/non-linear regimes has prevented direct measurement of the predicted narrow linewidths in high vacuum. For many levitated systems, the measured linewidths deviate from those predicted even at moderate vacuum in the 10$^{-4}$-10$^{-5}$~mbar range. The reduction of noise introduced by the levitating fields is a key challenge for achieving a stable oscillator. A fundamental limiting noise for optical levitation is the recoil of photons from the levitation laser itself~\cite{quidant_recoil}, while internal heating via absorption of laser light leads to motional heating~\cite{barker_nnano}. In contrast, a charged nanoparticle that is levitated using electrodynamic fields within a Paul trap, is an attractive levitated system as it is free from recoil induced heating~\cite{giacomo1,giacomo2} while only low light intensities are required for detection. In addition, large well depths in excess of 1 eV ($10^6$ K) allow operation of the trap in the harmonic regime, even for temperatures exceeding 300 K. We have recently demonstrated~\cite{trap_paper} that the charge on nanosphere in a Paul trap is stable over many weeks of measurement allowing stable oscillation frequencies limited only by the noise and drifts in the applied electric fields and environmental disturbances.

\begin{figure}[!ht]
\includegraphics[width=8cm]{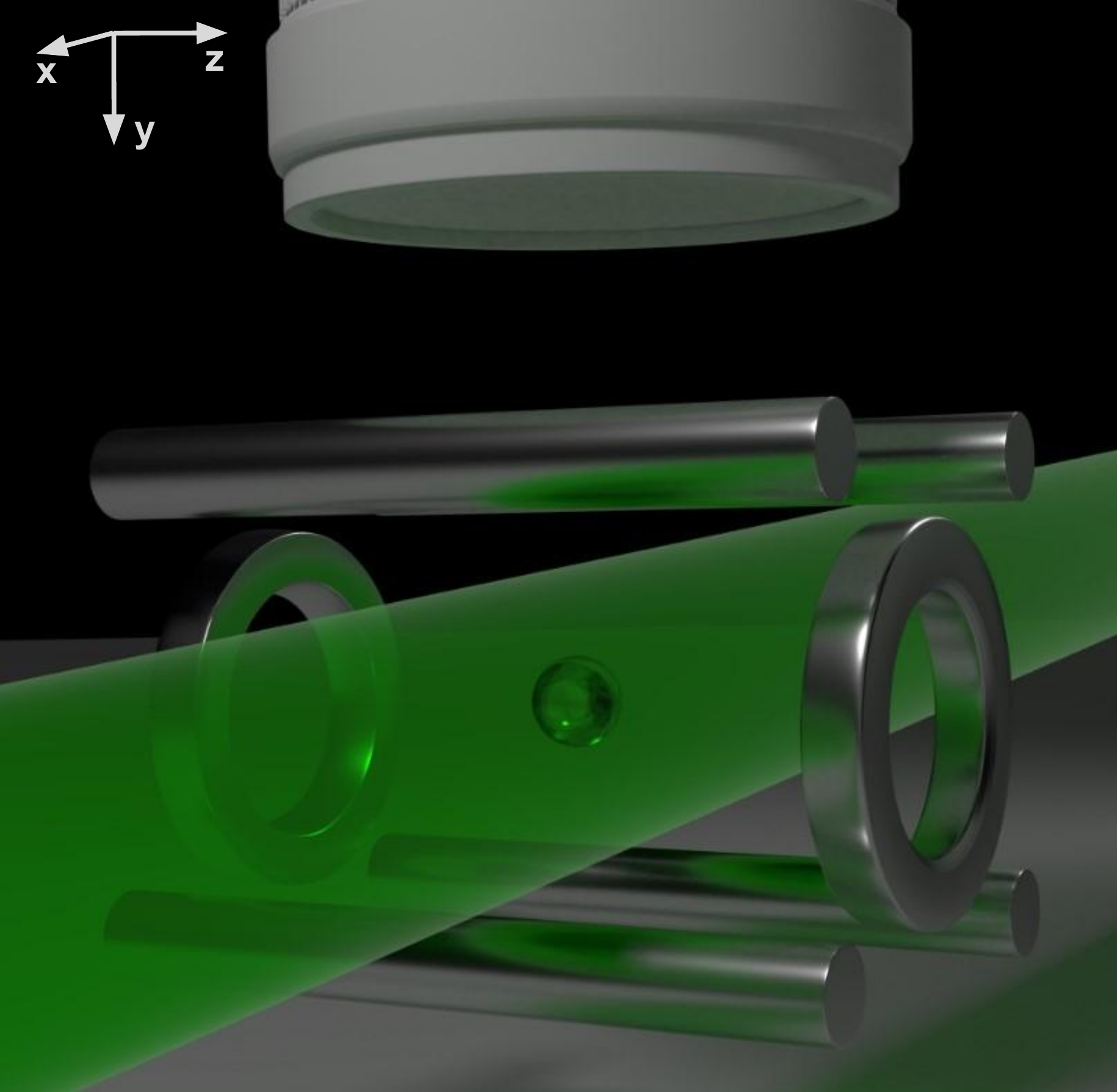}
\caption{Schematic view of the experiment. A silica nano-particle is levitated in a linear Paul trap at high vacuum. The motion is detected by illuminating the particle with a wide green laser field and imaging the light scattered at $90$\degree.}
\label{fig00}
\end{figure}

Here we report on a stable levitated nanoparticle oscillator operating at pressures down to 10$^{-7}$\,mbar in a room temperature environment. We outline the application of a phase sensitive detection method that allows us to remove the small residual variations in the trapping potential over the very long timescales required to measure an ultra-low linewidth of $81\pm\,23\,\mu$Hz.  This is to the best of our knowledge, the narrowest linewidth measured for a mechanical oscillator and paves the way for future precision experiments using levitated systems. We also characterise the important noise sources for this oscillator and outline a means to achieve even lower linewidth measurements and higher Q experiments for future experiments to test the macroscopic limits of quantum mechanics. Finally, we use our current oscillator to place new bounds on dissipative wavefunction collapse models.

We create our levitated nano-oscillator by trapping commercial silica nanospheres of radius 230 nm in high vacuum within a linear Paul trap. An illustration of the trap is shown in Fig.~\ref{fig00}. It consists of four stainless steel rods ($0.5$\,mm diameter) held in position by two supports made from printed circuit boards (PCB)~\cite{PCB} that are $7.0$\,mm apart. Two of the electrodes along one diagonal are grounded and the other two are driven by an AC electric field which provides a trapping potential in the plane perpendicular to the rods axis. To generate $3$D confinement, two additional ring end-cap electrodes, coaxial with the trap and printed directly on the PCB holders (not shown~\cite{trap_paper}), are kept at a DC constant voltage. The particles are injected in low vacuum ($\sim 10^{-1}$\,mbar) by means of electrospray ionisation.  A quadrupole mass filter~\cite{march} guides the particles from the input skimmer to the trapping region, which allows a better charge-to-mass ratio selection and an increased particle flux into the trap. The typical charge-to-mass ratio we levitate with this approach is between $0.05$ and $2$\,C/kg.

We exploit a simple detection scheme~\cite{imaging_paper}, where the trapped nanoparticle is illuminated with a $532$\,nm laser and the light scattered from it at $90$\degree\, is collected by a zoom objective lens and imaged onto a CMOS camera. The particle position is extracted by locating the coordinates of the brightest pixel. This simple approach allows real time acquisition of the particle motion in the plane defined by the field of view of the camera. The maximum sample rate ranges from $500$\,Hz to $800$\,Hz which is sufficient for resolving the low secular frequencies in the trap. The laser beam waist is $250 \mu$m located near the trap center with a typical laser power of $40$\,mW. The low intensity guarantees that the particle is not heated and that the internal temperature remains close to $300$\,K~\cite{barker_nnano}. Additionally, shot noise recoil heating from the probe beam is also negligible.

\begin{figure}[!ht]
\includegraphics[width=8.6cm]{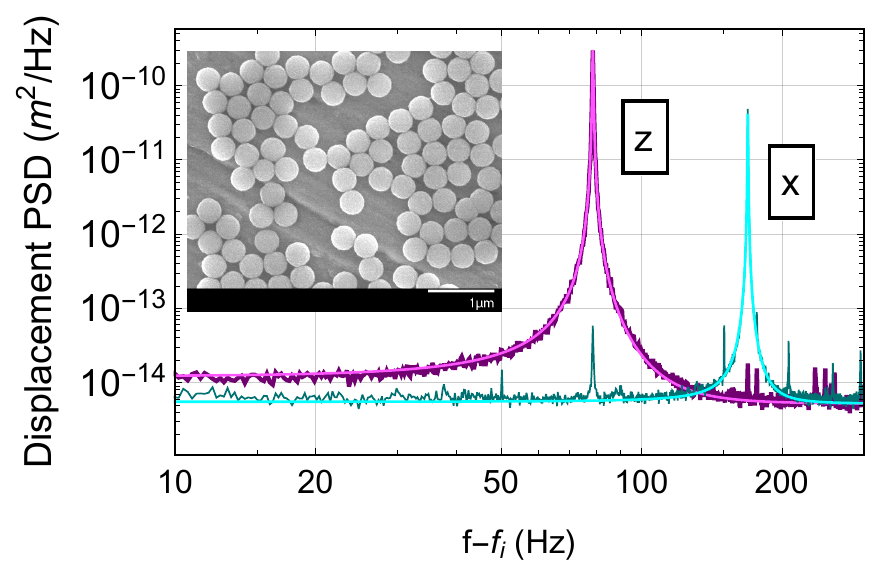}
\caption{Displacement PSD of a trapped nanoparticle along the two degrees of freedom monitored at a pressure $P_g=10^{-3}$\,mbar for a particle of $231$\,nm radius; also shown are fits with the mechanical susceptibility.  Inset: scanning electron microscope image of the nanoparticles used. }
\label{fig01}
\end{figure}

The oscillator dynamics are described by a set of Mathieu equations. The resulting trajectories are stable if the $a$ and $q$ parameters fall into the well known stability region~\cite{wineland}. Under the additional condition of $q\lesssim0.4$, a pseudo-potential approximation provides a very good description of the particle motion. In this case, the trapping potential is harmonic with frequencies along the $i$-axis given by $\omega_i\cong\omega_d/2\,\sqrt{a_i+q_i^2/2}$. Collisions with the residual gas result in a damped harmonic motion driven by a stochastic force, $F_{\text{th}}$, with power spectral density (PSD) $S_{F_{\text{th}}F_{\text{th}}}=2\,k_{B}T m\gamma$. An example of such thermal motion obtained with our detection scheme is shown in Fig.~\ref{fig01}. At the lowest pressure, other noise sources such as voltage noise on the trap electrodes or ambient displacement noise become important and must be controlled for stable operation.

One of the main advantages of the levitated platform lies in the possibility of achieving a very small coupling between the oscillator and the thermal bath. Assuming a particle internal temperature at equilibrium with the background gas at temperature T, inelastic collisions provide a damping rate given by $\gamma= 4\,\pi m_g R^2 v_t P_g/(3 k_B T m)(1+\pi/8)$\,\cite{epstein,Cavalleri2010}, where $v_t=\sqrt{8k_B T/(\pi m)}$ is the gas thermal velocity, $m_g$ the molecular mass and $P_g$ is the residual gas pressure. At relatively high pressures the damping rate can be accurately measured, even for large particles (i.e. $\sim1\-30\,\mu$m), however, as the pressure is reduced, it becomes increasingly difficult to perform accurate direct measurements. There are many examples in the literature demonstrating a saturation of the linewidth for levitated systems~\cite{Li2011,gieseler1}. Spectral estimation requires continuous monitoring for time scales much longer than the correlation time ($2/\gamma$) with the implicit requirement that the stability of the trap frequency is far better than the line width, i.e., $\delta\omega_i\ll\gamma$. On the other hand, for long correlation times, time-resolved techniques are usually preferred~\cite{nortup2019}. However, this approach requires driving the particle to large amplitudes in order to achieve a good signal-to-noise ratio which could lead to particle loss and, more often, to the exploration of highly non-linear regions of the trap potential.

We remove the problems related to the trapping potential stability by implementing a numerical phase sensitive detection. To this end, it is convenient to move to a frame rotating at the mechanical frequency $\omega_i$. The motion $u_i(t)$ along any axis, \textit{i}, can be decomposed into two quadratures $X_i(t)$ and $Y_i(t)$ according to $u_i(t)=X_i(t)\,cos{(\omega_i t)}+ Y_i(t)\,sin{(\omega_i t)}$.  To simplify the notation, we consider only one degree of freedom and denote its resonance frequency with $\omega_o$. If $\gamma\ll\omega_o$, the dynamical equations for the slowly varying quadratures are:
\begin{equation}\label{eq3}
\begin{split}
\dot{X}+\frac{\gamma}{2} X=\frac{1}{m \omega_o} f^{(1)}\\
\dot{Y}+\frac{\gamma}{2} Y=\frac{1}{m \omega_o} f^{(2)}
\end{split}
\end{equation}
\noindent where $f^{(k)}$ are stochastic force terms with autocorrelation functions $\langle f^{(k)}(t)f^{(j)}(t')\rangle= \delta_{kj}\delta(t-t') S_{FF}/2$, assuming a delta-correlated force noise $F$, with PSD $S_F$, driving the oscillator. The spectra of the two quadratures are then $S_{XX}(\omega)=S_{YY}(\omega)=\frac{S_{FF}}{2 (m w_o)^2}\frac{1}{\omega^2+\gamma^2/4}$. Experimentally, these PDSs are obtained by implementing a numerical lock-in amplifier, where we demodulate the displacement signal $u_o(t)$ at a frequency $w_{LO}$ and filter out the fast varying components at $2 \omega_o$. The two quadratures are still affected by changes of the trap frequencies or by a frequency difference between $\omega_o$ and the reference rotation $\omega_{LO}$; if $\omega_o=\omega_{LO}+\delta\omega$ their spectrum will be given by a low frequency Lorentzian peak centered at $\delta\omega$, of course, if it is a time dependent quantity the peak will be broadened and the linewidth estimation is incorrect or even impossible.

However, frequency drifts or offsets are removed if we look at the amplitude quadrature $R=\sqrt{X^2+Y^2}$. Although obtaining an analytical expression  for $S_{RR}(\omega)$ is not trivial, it is numerically straightforward. On the other hand, it is possible to readily evaluate the power spectral density (PSD) of $S_{R^2 R^2}(\omega)$ which is given by~\cite{autocorr}

\begin{equation}\label{eq4}
S_{R^2 R^2}(\omega)=\frac{8}{\gamma}\frac{1}{\omega^2+\gamma^2}\left(\frac{S_F}{2m^2 w_o^2}\right)^2.
\end{equation}

\noindent Importantly, the spectrum of $R^2$ is not affected by drifts or modulations of the resonance frequency as long as the rotating wave approximation remains valid and any change remains confined in the detection bandwidth defined by the low pass filters in the demodulation chain. 

\begin{figure}[!ht]
\includegraphics[width=8.6cm]{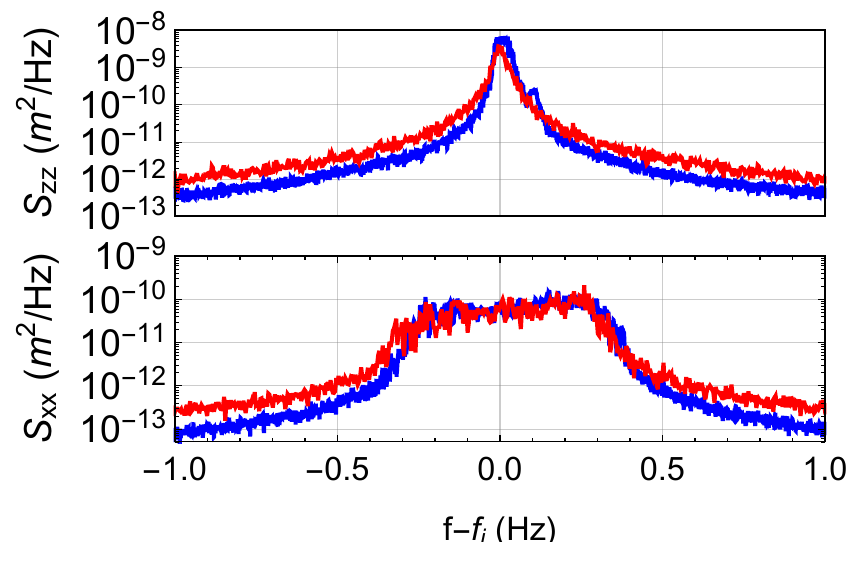}
\caption{Displacement PSD of a trapped nanoparticle along the two degrees of freedom monitored. The PSDs refer to two pressures $P_g=10^{-4}$\,mbar (red) and $P_g=2 \times 10^{-5}$\,mbar (blue). The linewidth measured exploiting the $R^2$ spectra are respectively $28.5\pm 0.7$\,mHz and $7.5\pm0.5$\,mHz.}
\label{fig1}
\end{figure}

Here we present data acquired by monitoring a single nanoparticle of mass $m= 9.6\pm0.4\pm0.9\times10^{-17}$\,kg~\cite{error} obtained by inducing charge jumps and measuring the resulting resonance frequency shift. Assuming a nominal density of $\rho=1850$\,kg/$m^3$ this corresponds to a radius of $r=231\pm7$\,nm.
\begin{figure*}[t]
\includegraphics[width=1\textwidth]{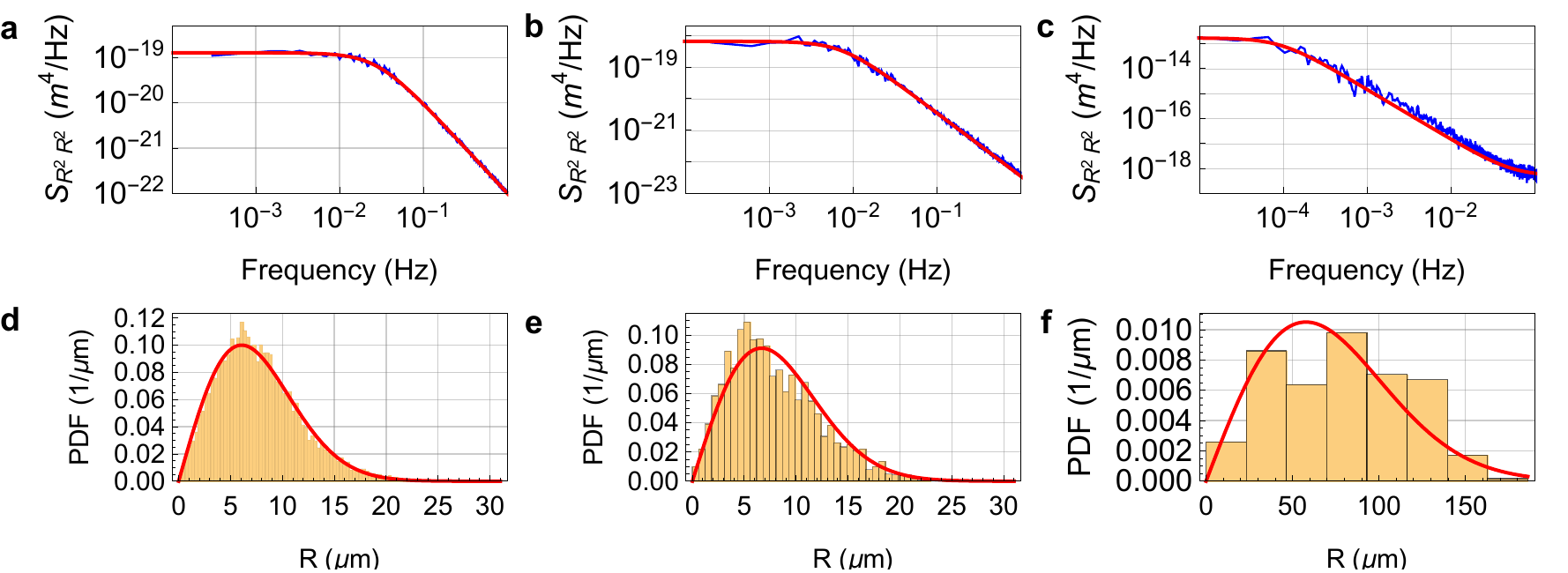}
\caption{Panels \textbf{a}, \textbf{b} and \textbf{c}: PSD  of the square amplitude quadrature for the motion along the \textit{x}-axis along with a fit following Eq.\,\ref{eq4} for different pressures $P_g=9 \times 10^{-5}$\,mbar, $2\times10^{-5}$\,mbar and $3 \times 10^{-7}$\,mbar respectively. Panels \textbf{d}, \textbf{e} and \textbf{f}: Probability density functions (PDF) of the amplitude quadrature $R$ for the same datasets along with the corresponding Rayleigh distribution. Each dataset corresponds to a continuous acquisition close to one day.
}
\label{fig2}
\end{figure*}
The frequency stability in our trap is of the order of $\delta\omega_{max}/\omega_o=0.003$ over a time scale of an hour. This is dominated by thermal drifts in the electronics that supply the electrode voltages and as well as to slowly changing stray fields. At the lowest pressures, this frequency drift is much larger than the expected line width. Furthermore, the behaviour along the two axis are significantly different. Along the trap axis (\textit{z}-axis in Fig.~\ref{fig00}) the resonance goes through a slow smooth drift, while in the orthogonal direction (\textit{x}-axis) there is an additional periodic modulation of $\delta \omega_x/2\pi=0.25$\,Hz with period of roughly an hour. This can clearly be seen in Fig.~\ref{fig1} where we show the displacement PSD along both axes at two different pressures ($P_g=10^{-4}$\,mbar (red curves) and $P_g=2\times10^{-5}$\,mbar (blue curves)). The effect of the frequency drift/modulation in the \textit{x}-axis is evident by the flattening of the spectral profile. Even in the \textit{z}-axis, where the PSD is not strongly perturbed, the conventional spectral approach yields a reliable line width measurement only at higher pressures.

\begin{figure}[!ht]
\includegraphics[width=8.6cm]{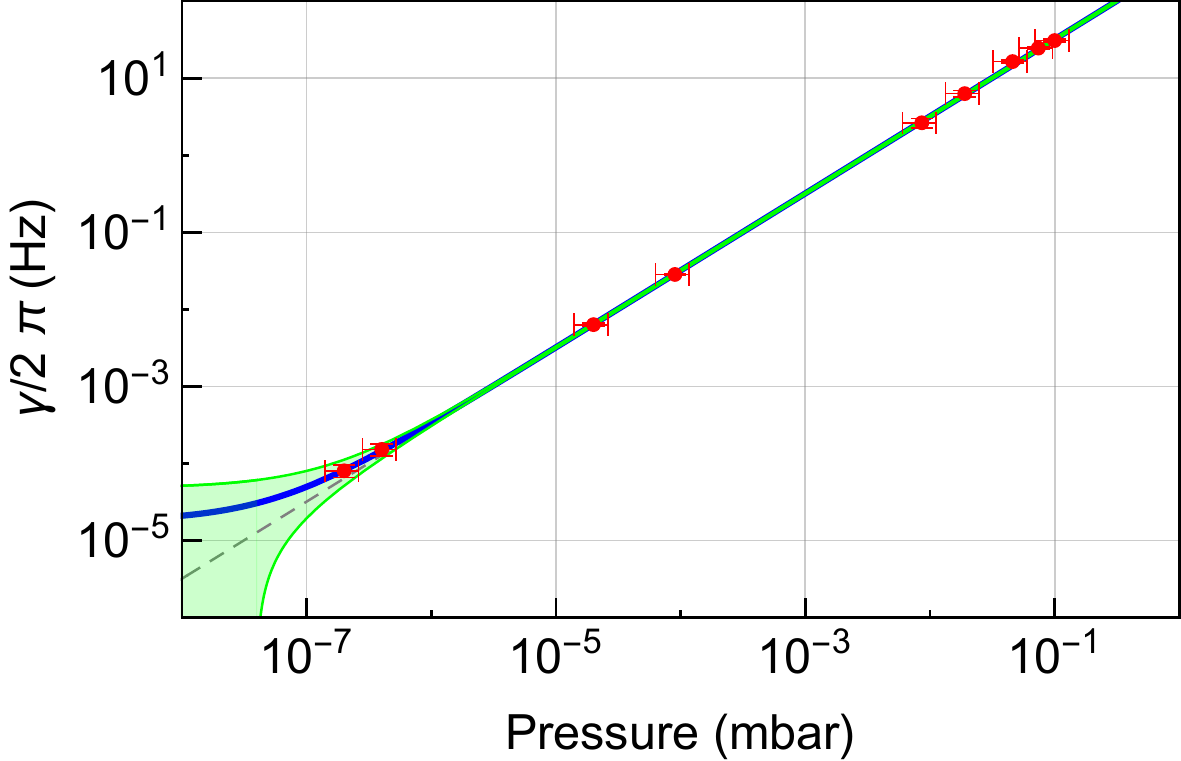}
\caption{Linewidths averaged over the two degrees of freedom measured and with a linear fit (blue) and $95$\% confidence bands (green). Also shown is the fit assuming no excess damping (dashed-grey).}
\label{fig3}
\end{figure}

In contrast, the linewidth can be determined from the PSD of $R^2$ as shown in Fig.~\ref{fig2} \textbf{a}, \textbf{b} and \textbf{c} for different pressures along the  \textit{x}-axis. Also shown is a fit to the data using Eq.~\ref{eq4}. The line width follows the expected behaviour from gas collisions very well despite the frequency modulation that was clearly visible in the conventional PSDs in Fig.~\ref{fig1}~\textbf{b}. We point out that the modulation in the secular frequency $\delta \omega_x$, at the lowest pressure, is 2500 times larger than the line width $\gamma$, illustrating the utility of this method for extracting narrow linewidths in the presence of slow, large modulation and drifts in the trap frequency. To test the accuracy of this approach we fit the much wider high pressure measurements performed with the standard spectral method and compare them to the linewidths obtained from the $R^2$ spectra. We find perfect consistency between the two methods (see Methods).  The linewidths as a function of pressure, averaged between the two degrees of freedom, are fitted  with a simple line, i.e., $\gamma=\gamma_{\text{exc}} + k\,P_g$, to allow the estimation of a possible excess damping $\gamma_{\text{exc}}$. The fit is shown in Fig.~\ref{fig3} along with the $95$\% confidence bands. With the same confidence level we find $\gamma_{exc}=18\pm30\,\mu$Hz which is consistent with zero confirming that only gas damping is affecting the particle dynamic. The minimum measured linewidth is the ultra-low value of $81\pm\,23\,\mu$Hz.

We show in Fig.~\ref{fig2} \textbf{d}, \textbf{e} and \textbf{f} histograms of the amplitude quadrature at different pressures. This is the well known Rayleigh distribution $\mu(r;\sigma_i)=r/\sigma_i^2\, e^{-r^2/\sigma_i^2}$ with mean and variance given by $\langle r\rangle=\sqrt{\pi/2}\,\sigma_i$ and $\sigma_{\mu}^2=(4-\pi) \sigma_i^2/2$, respectively. If the motion along the $i$-axis is in thermal equilibrium with the environment we simply have $\sigma_i^2=k_B T/m \omega_i^2$ which is the usual thermal variance for an harmonic oscillator. The three data sets have a similar total observation time of the order of one day. This implies that the number of statistically independent points significantly drops when the pressure is reduced to the minimum value. The sample size goes from approximately $10000$ at $P_g=10^{-4}$\,mbar to $100$ at $P_g=3\times10^{-7}$\,mbar as can clearly be observed in Fig.~\ref{fig2}~\textbf{d} and \textbf{f}. We exploit the properties of the Rayleigh distribution to verify that the sample size is big enough to be representative of the distribution. The first two moments allow an estimation of the motional variance which can be compared with each other. Specifically, at $P_g=10^{-4}$\,mbar we have $\Delta\sigma_i/\langle\sigma_i\rangle=0.005$ while at $P_g=3\times10^{-7}$\,mbar we find $\Delta\sigma_i/\langle\sigma_i\rangle=0.08$ which is significantly bigger but perfectly in line with what is expected from the sample size.

\begin{figure}[!ht]
\includegraphics[width=8.6cm]{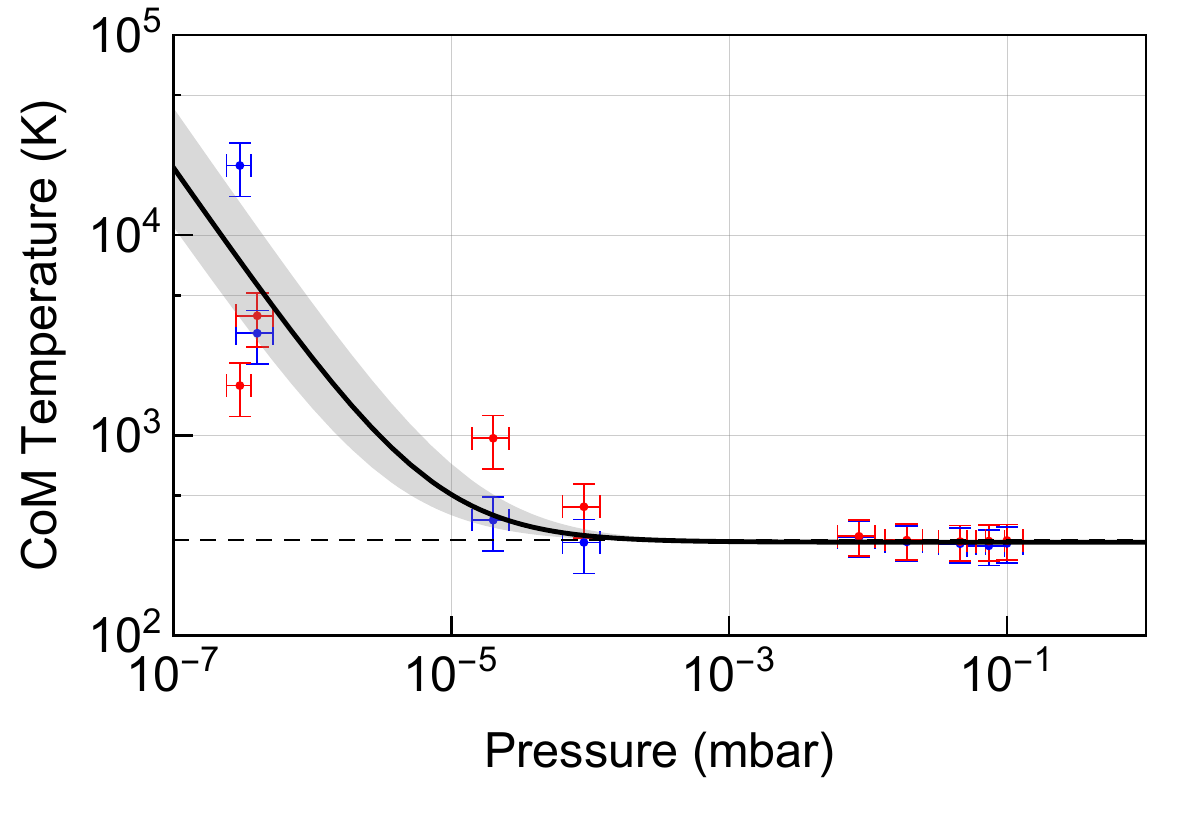}
\caption{Centre of mass (CoM) effective temperature as a function of pressure along both \textit{x}-axis (blue) and \textit{z}-axis (red); the average between the two degrees of freedom is fitted assuming a $1/P_g$ contribution that would rise from a white non-thermal force noise. A $3$\,dB temperature increase occurs at $P_g=7.3\times10^{-6}$\,mbar. Dashed gray line marks the measured room temperature of $293$\,K.}
\label{fig4}
\end{figure}

It is clear, from both the PSD spectra and the Rayleigh distributions of Fig.~\ref{fig2}, that the variance of the stochastic motion increases as the pressure is reduced. This behaviour is also shown as an increase in the effective temperature as the pressure is reduced in Fig.~\ref{fig4}. Here we plot the effective temperature along both trap axes as a function of pressure. This calibration does not rely on the assumption of thermal equilibrium at room temperature but exploits the absolute calibration available from using the camera\cite{imaging_paper}. For the high pressures ($P_g>\,10^{-3}$\,mbar) the mean effective temperature is $\langle T\rangle=293\pm3\pm25$\,K, where the systematic error comes from the uncertainty in the mass measurement. The excess force noise leads to an increase in effective temperature that varies inversely with pressure. A fit to the average temperature on the two axes gives $T_{\text{eff}}= 293 (1+7.3\times10^{-6}/P_g)$. We obtain a value for the excess force noise from the PSD of $S_{ff}\simeq 1\times10^{-38}$\,N$^2/$Hz (heating rate of $240$\,MHz). For the nanoparticle illuminated by the detection beam, the back action due to photon recoil is approximately $4\times10^{-43}$\,N$^2/$Hz ($9$\,kHz) which is insignificant compared to the thermal noise of $10^{-40}$\,N$^2/$Hz  ($3$\,MHz) at $10^{-7}$\,mbar. Assuming the excess noise that we observe is dominated by voltage noise we estimate its value to be $\sqrt{S_{VV}}\simeq10\,\mu$V$/\sqrt{\mathrm{Hz}}$ which is consistent with the measured voltage noise. This is determined from $S_{ff}=(n_{\text{ch}}\,e)^2 S_{VV}/D^2$, where $D=2.3$\,mm is the characteristic distance related to the AC electrode geometry (see Methods) and $n_{\text{ch}}\simeq80$ is the number of charges on the nanoparticle.

Our nanoscale oscillator, with a Q of 1.5$\times$10$^6$, compares well with the highest Q-factors ever reported for relatively low frequency oscillators, particularly for operation at room temperature. The most notable are balanced torsional oscillators (i.e., QPO)~\cite{QPO1,QPO2} where Qs of almost $10^6$ are reached for higher oscillation frequencies of a few kHz. A single-crystal silicon oscillator has demonstrated a linewidth of $\sim500\,\mu$ Hz at $300$\,K~\cite{McGuigan1978}. For all these systems, however, only a very limited number of normal modes have such high Q-factors.  This is in stark contrast with the levitated case described here where all three translational degrees of freedom of the oscillator experience the same dissipative forces, while the internal modes are completely decoupled as the lowest modes have frequencies greater than GHz~\cite{Wheaton2014}.

We now describe the use of our narrow linewidth oscillator, coupled with our ability to predict the simple variation of damping with pressure, to put new limits on two dissipative models for wavefunction collapse. This includes the dissipative continuous spontaneous localisation (CSL)~\cite{GRW} and the dissipative Di\'{o}si-Penrose (DP)~\cite{Penrose1996,Penrose1998,diosi} models. Recently, a microscale oscillator cooled to $20$\,mK has been used to place stringent constraints on the CSL model~\cite{Vinante1,vinante2}. Its effect on an oscillator is an additional stochastic force noise which has a white spectral density $S_{\text{CSL}}=\hbar^2 \eta_{\text{CSL}}$ that induces a collapse of the wavefunction. This is characterised by a collapse strength $\eta_{\text{CSL}}$ (see methods) which depends on the geometry and density of the oscillator and upon two parameters, namely, the collapse rate $\lambda$ and a characteristic length $r_C$. The conventional CSL model is not energy conserving and a dissipative version (dCSL)\,\cite{Smirne2015} has been proposed which removes the energy divergence due to the standard CSL force noise. The additional dissipation determines a finite equilibrium temperature for any given system and an additional parameter $T_{\text{dCSL}}$ is introduced which characterises the temperature of the collapse noise. In the dCSL framework, the force noise is no longer white and for a nanoparticle it has a spectral density~\cite{PRA_dCSL} $S_{\text{dCSL}}(\omega)=\hbar^2 \eta_{\text{dCSL}} [1+\kappa^2 m^2 (\gamma_t^2+\omega^2)]$ where $\gamma_t=\gamma + \gamma_{\text{dCSL}}$ is the total damping rate and $\kappa=\gamma_{\text{dCSL}}/2\hbar\eta_{\text{dCSL}}$. In the limit of an infinite temperature the standard CSL is recovered.

\begin{figure*}[!t]
\includegraphics[width=1\textwidth]{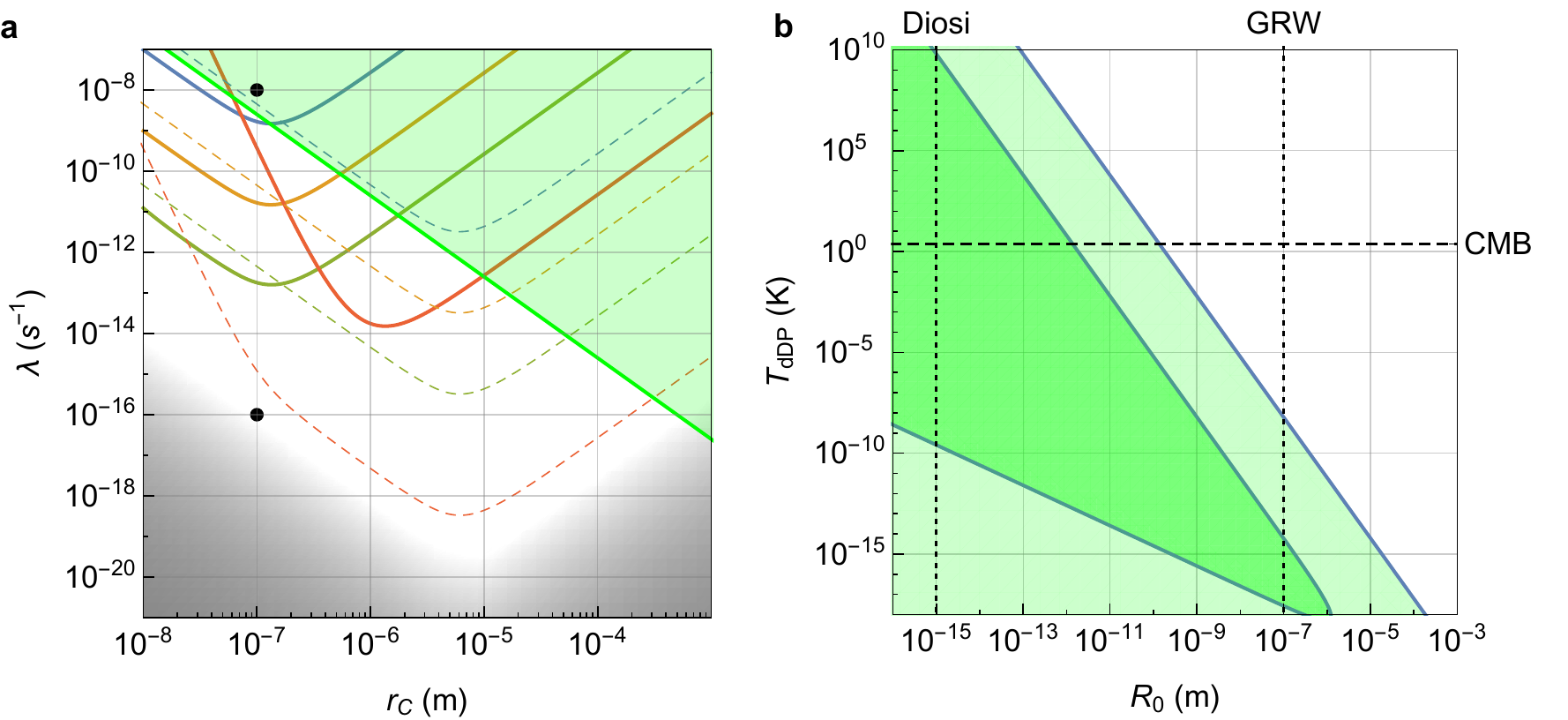}
\caption{Experimental bounds on dissipative collapse models. \textbf{a} Exclusion region for the dCSL parameters $\lambda$ and $r_C$. Continuous lines: upper bounds from current line width measurements for a $231$\,nm  radius silica particle at $3\times 10^{-7}$\,mbar for different values of $T_{\text{dCSL}}$: $1$~K (blue), $10^{-2}$~K (yellow), $10^{-4}$~K (green) and $10^{-7}$~K (red). The dashed lines are the expected improved bounds from future measurements on a $10\,\mu$m radius silica particle; same color code for $T_{\text{CSL}}$. The green region corresponds to upper bound from LISA pathfinder~\cite{pathfinder} and the grey region is the lower bound from theoretical arguments\,\cite{Toros2017,torovs2018bounds}. The black markers refer to the Adler~\cite{Adler_1,Adler_2} (top) and GRW~\cite{GRW} (bottom) values. \textbf{b} Exclusion region for the single-particle dDP model. The darker green corresponds to the bounds from the current experiment. The lighter green region is the expected improved bounds from future measurements on a $10\,\mu$m radius silica particle. We chose to limit the temperature to the lowest conjectured value of $10^{-18}$~K associated with Hawking radiation from a supermassive black hole~\cite{HAWKING1974,hawking1975}. CMB marks the cosmic microwave background temperature.}
\label{fig5}
\end{figure*}

The DP model has a structure quite similar to the CSL with the main difference being that the scale of the collapse strength is set by the gravitational constant $G$, thus introducing a stronger connection to gravity for the collapse mechanism. The standard DP model is also not energy conserving, and a dissipative version has been proposed~\cite{dDP}. Since the collapse strength is fixed, the dissipative DP (dDP) model is characterised by only two parameters, namely, the temperature of the collapse noise $T_{\text{dDP}}$ and a cut off distance $R_0$ which plays a similar role as $r_C$ in the dCSL model.

The CSL and DP family of models usually discussed in the literature predict that the collapse strength for the centre-of-mass dynamics depends on the intensive and extensive properties of the bulk material, such as composition and shape. This is the version we consider for dCSL. On the other hand, we can also consider the dDP model in the single-particle version, which depends only on the total mass of the system, but not on any other property of the system (see methods). In the following, we consider this latter version of the dDP model, while our experiment does not put significant bounds on the former one.

For both the dCSL and dDP models, the force noise sensitivity required to place significant bounds is quite far from what can be achieved in our current experiment. However, we can exploit the extremely narrow linewidth of our oscillator to put significant constraints on these models. For the levitated oscillator the damping rate is well defined by the collisions with the surrounding gas and has a linear dependence with pressure. This allows us to extrapolate its value at zero pressure with any offset potentially due to the collapse process. We find an upper limit for the damping rate due to the collapse mechanisms of $\gamma_{\text{cm}}\leq48\mu$Hz at the 95\% confidence level which is compared to the predictions from the models.

Exploiting this upper limit we derive new bounds for dCSL which are shown in Fig.~\ref{fig5}\textbf{a}. These new bounds assume different dCSL temperatures. When $T_{dCSL}=1$~K our bounds are comparable to existing ones~\cite{pathfinder}. However, when $T_\text{dCSL}$ is reduced to $10^{-7}$~K~\cite{PRA_dCSL} the excluded parameter space is significantly increased reaching a minimum collapse $\lambda\simeq10^{-14}$~s$^{-1}$ for an $r_C=1.5\,\mu$m. Remarkably, for these values the force noise sensitivity required to place an equivalent bound on both CSL and dCSL is $\sim 10^{-50}$~N$^2/$Hz which would be achievable only for a thermal particle at a pressure of $\sim10^{-15}$\,mbar (assuming the same particle size).

The excluded parameter space for the dDP model in the single-particle version is shown in Fig.~\ref{fig5}\textbf{b}. The value for $R_0$ originally proposed by Di\'{o}si is excluded on an extremely wide temperature range from $10^{-10}$~K to $10^{10}$~K while at the GRW value the exclusion is limited to very unrealistic $T_{\text{DP}}$ values. A value considered reasonable is the cosmic microwave background (CMB) temperature~\cite{PRA_dCSL}; in this case the excluded $R_0$ goes from $10^{-18}$~m to $10^{-12}$~m. Interestingly, the excluded region has no upper bound and extends to $T_{\infty}$ for increasingly smaller values of $R_0$.

In conclusion we have demonstrated that a narrow linewidth, high Q, levitated nano-oscillator can be created by trapping nanoparticles in the Paul trap. We show that, even in the presence of voltage noise from the electronics, the trap is stable enough to measure a record low mechanical linewidth of $\sim80~\mu$Hz at room temperature. This compares well with other high-Q low frequency oscillators that typically achieve similar performances at cryogenic temperatures. Using the narrow linewidth of this oscillator we place new constraints on two dissipative models for wavefunction collapse, namely the continuous spontaneous localisation and the Di\'{o}si-Penrose in the single-particle version. By utilising lower noise electronics, and a larger mass oscillator which can be incorporated in this system, we expect to be able to provide even more stringent limits on standard and dissipative collapse models in the future. Considering a $10\,\mu$m radius silica particle, and measuring a linewidth an order of magnitude smaller than measured here, would have a significant impact. In the case of the dCSL it would allow us to almost entirely exclude the $\lambda$-$r_C$ parameters subspace for the lowest $T_{\text{dCSL}}$~\cite{Toros2017,torovs2018bounds}. In the case of the dDP, we could exclude $R_0$ values almost two orders of magnitude larger at the CMB temperature. Finally, if such a particle is thermal at $P_g\sim10^{-10}$\,mbar our system would allow us to probe the standard CSL for collapse rates approaching $10^{-13}$\,s$^{-1}$ at a characteristic length of $\sim6\,\mu$m thus extending well beyond the current bounds.

\section{Methods}

\subsection{Trap and detection}

The linear Paul trap, described in the main text, is characterised by $4$ parameters, two of which relate to the actual trap geometry and two to the efficiency coefficients that quantify non-perfect quadratic potentials. The former are the distance between the centre of the trap and the AC and DC electrodes, respectively where $r_o=1.1$~mm and $z_o=3.5$~mm. The latter are $\eta=0.82$ which represents the quadrupolar coefficient of the multipole expansion of the AC pseudo-potential and $\kappa=0.086$ which is the quadratic coefficient of the DC potential at the trap centre~\cite{efficiency1,efficiency2}. These are calculated with numerical simulations based on the finite element method (FEM). The Mathieu stability parameters are given by:

\begin{equation}\label{eq1sup}
\begin{split}
a_{x}=a_{y}=-\frac{1}{2}a_{z}=-\frac{q}{m} \frac{4\kappa\,U_{o}}{z_{o}^{2}\,\omega_{d}^{2}} \\
q_{x}=-q_{y}=\frac{q}{m}\frac{2\eta V_{o}}{r_{o}^{2}\,\omega_{d}^{2}},\,q_{z}=0
\end{split}
\end{equation}

\noindent where $U_o$ and $V_{o}$ are respectively the DC potential and the amplitude of the AC field, $q$ is the total electric charge and $m$ is the mass of the particle. Typical values for the DC voltage are from $50$ to $150$~V, while for the AC we have amplitudes between $100$ and $300$\,V with frequencies from $1$ to $5$\,kHz. A detailed characterisation of the trap, its stability and the loading approach can be found in Ref.~\cite{trap_paper}. Finally, additional information on the detection scheme implemented and its potential improvements can be found in Ref.~\cite{imaging_paper}.

\subsection{Effect of voltage noise}

The most important contribution to the force noise acting on the nanoparticle is voltage noise from the AC electrodes. Its evaluation requires the knowledge of the electric field near the trap centre which was obtained from finite element modelling simulations based on our trap voltages and geometry. With a voltage $V$ applied on both AC electrodes, and taking contributions up to the first order, we obtain

\begin{equation}\label{eq_field}
\frac{E(x,y)}{V}=\left(\frac{1}{D}+\frac{<x>}{D_1}\right)\,\check{x}+\left(\frac{1}{D}+\frac{<y>}{D_1}\right)\,\check{y}
\end{equation}

\noindent where $<x>$ and $<y>$ are the mean particle positions, $D=2.3$~mm and $D_1=1.5\,\mu$m. Assuming uncorrelated voltage noise and that the particle is at the trap centre, the force noise PSD can then be evaluated by $S_{ff}=(n_{cn}\,e)^2 S_{VV}/D^2$~\cite{voltage_noise} which gives a voltage noise of $\sqrt{S_{VV}}\simeq10\,\mu$V$/\sqrt{\mathrm{Hz}}$, as in the main text. This value, however, is a rough estimation for several reasons: first, it assumes the particle is perfectly positioned in the trap centre and variations of the the mean position over the long experimental times could lead to deviations of the measured temperatures at low pressure (Fig.~\ref{fig4}). Secondly, from geometrical considerations, the force noise on the particle due to voltage noise on the electrodes should significantly lower along the \textit{z}-axis (endcaps); this is clearly not the case and suggests a significant cross-coupling between degrees of freedom likely to be due to non-linearity in the potential. Finally, since the thermal induced variance along $x$ and $y$ is greater than $D_1$, voltage noise is actually a multiplicative noise and a detailed would require a more sophisticated model.

\subsection{Collapse rates}

In the dCSL model the collapse strength, $\eta_{\text{dCSL}}$, for a spherical particle of radius $r$ and homogeneous density $\rho$ is given by~\cite{PRA_dCSL}:

\begin{alignat}{1}
\eta_{\text{dCSL}}= & \frac{3\lambda r_{C}^{2}m^{2}}{\left(1+\chi_{\text{C}}\right)r^{4}m_{0}^{2}}\bigg[1-\frac{2r_{C}^{2}\left(1+\chi_{\text{C}}\right)^{2}}{r^{2}}\nonumber \\
 & +e^{-\frac{r^{2}}{r_{C}^{2}\left(1+\chi_{\text{C}}\right)^{2}}}\left(1+\frac{2r_{C}^{2}\left(1+\chi_{\text{\text{C}}}\right)^{2}}{r^{2}}\right)\bigg],\label{eq:dCSL}
\end{alignat}
where $m$ is the total mass of the system, $m_{0}$ is the nucleon mass, $\chi_{\text{C}}=\hbar^2/(8m_{a}k_{B}T_{\text{dCSL}}r_{C}^{2})$,
and $T_{\text{dCSL}}$ is the thermalisation temperature. In the limit of no dissipation, $T_{\text{dCSL}}\rightarrow\infty$, we have $\chi_{\text{C}}\rightarrow0$,
and thus one recovers the collapse strength of the standard CSL model, $\eta_{\text{dCSL}}\rightarrow\eta_{\text{CSL}}$. On the other hand, in case of strong dissipation when $a\ll r_{C}(1+\chi_{\text{C}})$, we obtain an approximate expression
\begin{equation}
\eta_{\text{dCSL}}=\frac{mm_{a}\lambda r_{C}}{2a^{3}m{}_{0}^{2}(1+\chi_{\text{C}})^{2}}\min\left[1,\frac{r^{3}}{r_{C}^{3}(1+\chi_{\text{C}})^{3}}\right]
\end{equation}
where $m_{a}$ is the average mass of a nucleus ($\simeq$ average atomic mass), and $a$ is the lattice constant~\cite{nimmrichter2014optomechanical}. Finally, the dissipation rate is given by $\gamma_{\text{dCSL}}=\eta_{\text{dCSL}}4r_{C}^{2}\chi_{\text{C}}(1+\chi_{\text{C}})m_{a}/m$.

The calculation for the dDP model under the assumption of a homogeneous sphere can be carried out similarly as for the dCSL model. Specifically, we find the following collapse strength:

\begin{alignat}{1}
\eta_{\text{dDP}}= & \frac{Gm_{a}^{2}}{\sqrt{\pi}r^{6}\hbar}\bigg[\sqrt{\pi}r^{3}\text{Erf}\left(\frac{r}{R_{0}(1+\chi_{\text{0}})}\right)\nonumber \\
 & +(1+\chi_{\text{0}})R_{0}\bigg\{ r^{2}\left(e^{-\frac{r^{2}}{(1+\chi_{\text{0}})^{2}R_{0}^{2}}}-3\right)\nonumber \\
 & +2(1+\chi_{\text{0}})^{2}R_{0}^{2}\left(1-e^{-\frac{r^{2}}{(1+\chi_{\text{0}})^{2}R_{0}^{2}}}\right)\bigg\}\bigg]\label{eq:dDP}
\end{alignat}
where $\chi_{\text{0}}=\hbar^2/(8m_{a}k_{B}T_{\text{dDP}}R_{0}^{2})$, and $T_{\text{dDP}}$ is the thermalisation temperature. In the limit of no dissipation, $T_{\text{dDP}}\rightarrow\infty$, we have $\chi_{\text{dDP}}\rightarrow0$, and the collapse strength of the standard DP model is recovered, $\eta_{\text{dDP}}\rightarrow\eta_{\text{DP}}$ ~\cite{nimmrichter2014optomechanical}. Using a lattice model for the crystal structure we obtain an approximate formula for the case of strong dissipation, $a\ll R_{0}(1+\chi_{\text{0}})$, given by

\begin{equation}
\eta_{\text{dDP}}=\frac{Gmm_{a}}{6\sqrt{\pi}a^{3}\hbar}\min\left[1,\frac{r^{3}}{R_{0}^{3}(1+\chi_{\text{0}})^{3}}\right]
\end{equation}
The damping rate for the dDP model is given by $\gamma_{\text{dDP}}=\eta_{\text{dDP}}4R_{0}^{2}\chi_{0}(1+\chi_{0})m_{a}/m$.

The CSL and DP family of models share many common features, among which the amplification of the collapse strength with the size of the system~\cite{bahrami2014role}. Such an amplification mechanism is supposed to accomplish a dual task: on the one hand, agreement with the quantum mechanical predictions for very small systems, and that classicality emerges for large systems~\cite{nimmrichter2013macroscopicity}.
The amplification mechanism has its origin in microscopic derivations where the centre-of-mass dynamics of a bulk material is obtained starting from the dynamics of nucleons or nuclei. In the present case this is reflected in geometrical factors $\eta_{\text{dCSL}}$ and $\eta_{\text{dDP}}$ that describe the effect of the intensive and extensive properties of the system on the centre-of-mass dynamics. In the above formulae
we have chosen the nuclei as the building blocks of the models, motivated by considerations about gravity in the Newtonian regime~\cite{tilloy2017principle}.

However, one could also postulate that the centre-of-mass motion is always in the single-particle form. In particular, one can require that the centre-of-mass dynamics depends only on the total mass $m$ of the system, but not on other intensive or extensive properties of the bulk system. This latter choice is suggestive of a hypothetical underlying theory of spontaneous collapse models, which includes notions
of Einstein's General relativity~\cite{penrose2014gravitization}. In this latter case, one has that the collapse strengths are given by $\eta_{\text{dCSL}}=\frac{\lambda m^{2}}{2m_{0}^{2}r_{C}^{2}(1+\chi_{\text{C}})^{5}}$ and $\eta_{\text{dDP}}=\frac{Gm^{2}}{6\sqrt{\pi}\hbar R_{0}^{3}(1+\chi_{\text{0}})^{3}}$, and the dissipation rates are obtained by replacing $m_{a}$ with $m$ and $a$ with $r$ in the corresponding formulae.

\subsection{Linewidth measurement}

\begin{figure}[!ht]
\includegraphics[width=8.6cm]{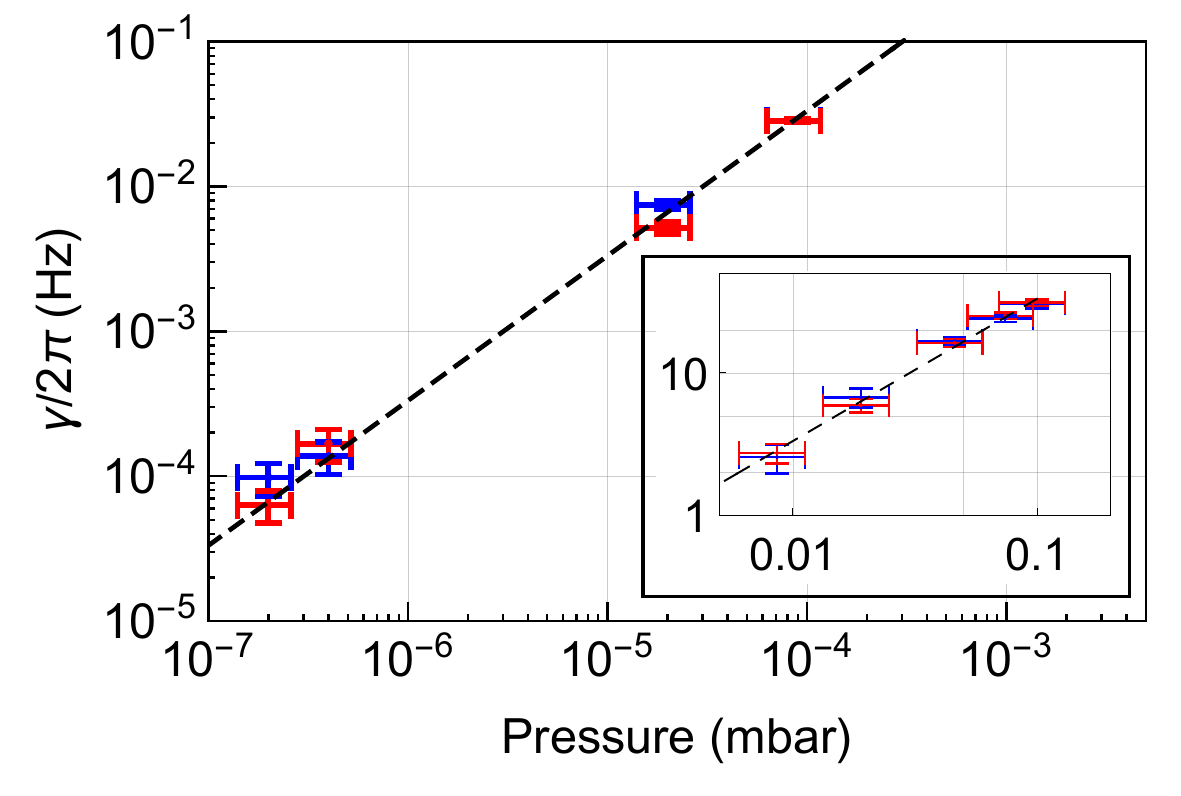}
\caption{Measured linewidths as a function of pressure along both \textit{x}-axis (blue) and \textit{z}-axis (red) obtained from the $R^2$ spectra. Inset: linewidth measured at high pressure by standard spectral analysis. The dashed line is a fit of the expected $1/P_g$ behavior. Only the data in the inset have been used for the fit.}
\label{fig3s}
\end{figure}

In order to verify the accuracy of the $R^2$ method to measure the linewidth we compare the low pressure data with the fit obtained exclusively with the high pressure measurements performed with the standard spectral method using the PSD. This is shown in Fig.~\ref{fig3s} where the line widths along the two monitored axes are fitted separately. All the low pressure measurements are consistent with the expected behaviour.

\section{acknowledgments}
The authors acknowledge funding from the EPSRC Grant No. EP/N031105/1 and the H2020-EU.1.2.1 TEQ project Grant agreement ID: 766900. AP has received funding from the European Union’s Horizon 2020 research and innovation programme under the Marie Sklodowska-Curie Grant Agreement ID: 749709 a N.P.B. acknowledges funding from the EPSRC Grant No. EP/L015242/1.

\bibliographystyle{naturemag}
\bibliography{paper_linewidth_dCSL}

\begin{thebibliography}{10}
\bibitem{error}
 \bibinfo{note}{The first error is statistical, the second systematic.}
\end{thebibliography}

\end{document}